\documentstyle[prl,aps,multicol]{revtex}
\begin{document}

\title{Asymptotic dynamics of short-waves in nonlinear 
        dispersive models }
\author{ M. A. Manna and V. Merle \\
{\em Physique Math\'ematique et Th\'eorique, CNRS - UM2,
34095 MONTPELLIER (France)}}
\maketitle

\begin{abstract}
The multiple-scale perturbation theory, well known for long-waves, is
extended to the study of the far-field behaviour of {\it short-waves},
commonly called ripples. It is proved that the
Benjamin-Bona-Mahony-Peregrine equation can propagates short-waves.
 This result
contradict the Benjamin hypothesis that short-waves tends not to
propagate in this model and close a part of the old controversy between
Korteweg-de Vries and Benjamin-Bona-Mahony-Peregrine equations. We
shown that a nonlinear (quadratic) Klein-Gordon type equation
substitutes in a short-wave analysis the ubiquitous Korteweg-de Vries
equation of long-wave approach. Moreover the kink solutions of 
 $\phi^{4}$ and sine-Gordon equations are understood as an all orders
asymptotic behaviour of short-waves. It is proved that the antikink
solution of  $\phi^{4}$ model which was never obtained perturbatively
can be obtained by perturbation expansion in the wave-number $k$ in
the short-wave limit.

\end{abstract}

\begin{multicols}{2}

\paragraph*{Introduction.}

The method of multiples scales (or reductive perturbation method)
is a very powerfull method
to study a large number of physical phenomena, 
in particular wave motion in nonlinear dispersive systems.  
It is well known that the far field dynamics of a
long-wave (LW) with a small amplitude in a nonlinear and dispersive
system can be reduced almost always to a reduced set of model equations
such as Boussinesq, Korteweg-de Vries (KdV), Modified KdV, etc \cite
{Taniuti,su,kodama,jeffrey}.

The purpose of this work is to look for the far field behaviour of
{\it short-waves} (SW) in some nonlinear and dispersive systems. In
order to see how SW propagate we reformulated the method of multiples
scales. The systems we will consider are three. Firstly a system coming
from hydrodynamic: The Benjamin-Bona-Mahony-Peregrine (BBMP)
equation. Secondly we will study SW in two important classical 
relativistic field
theory models: The {\it $\phi^{4}$} model ($\phi^{4}$) and the
sine-Gordon (SG) equation.

We will prove for BBMP (\ref{bbmp}) that SW can build up
the same soliton solution as obtained from LW\cite{kraenkel1}, 
hence rising the question
of the unicity of the soliton description.
We will prove that the antikink (or kink) solution of $\phi^4$ model
(\ref{phi4}) which {\it can
not be obtained} as a perturbative solution in the non-linearity
paramenter $\lambda$ occurs as a perturbative solution in the
wave-number $k$ in the SW limit.
Moreover the kink solution of sine-Gordon 
(\ref{sg}) does not enter the classical LW perturbation scheme. We
will prove that it appears as a perturbative solution for SW.

\paragraph*{Short-Wave Approach.} 
We are interested in the problem of the asymptotic dynamics of SW in
nonlinear and dispersive systems.
To perform this study,
all degrees of dispersion of the system are taken into account in a
Taylor
expansion of the linear dispersion relation $\omega(k)$ around a
large value of the wave-number $k$.
The asymptotic dynamics of SW for $t\rightarrow\infty$ is considered 
via
the introduction of an infinite number of {\it slow time} variables 
$\tau_{1},\tau_{3},\tau_{5},...$  and a {\it fast space} variable
 $\zeta$ 
following the extension theory of Sandri \cite{sandri}. The use
of this fast variable and an infinite series of slow time variables
constitute the first key of the SW approach.

The solution is expanded in the form of a power series in a small
parameter $\epsilon$ proportional to the inverse of the wave-number
$k$. The
perturbative series solution is secular. It is regularized (uniform
expansion), through a renormalization of 
the frequency. This results to the celebrated Stokes' hypothesis  on
frequency-amplitude dependence in water waves \cite{Whitham}. 

The Stokes' hypothesis is actually the second key tool of our approach.
Indeed, for LW asymptotic description, the KdV
 \cite{kraenkel,pereira}
 or MKdV \cite{manna} hierarchies
occult the need of this tool, as they naturally provide the correct 
series expansion of the frequency.

\paragraph*{Basic Models.}

Hence the problem is the asymptotic
behaviour of a SW in Benjamin-Bona-Mahony-Peregrine
(BBMP) \cite{Bona}
and in two classical relativistic nonlinear models: $\phi^{4}$ 
and sine-Gordon (SG) \cite{rajara}:
\begin{eqnarray}\label{bbmp}
({\rm BBMP})\ & u_{t} + u_{x} - u_{xxt} &=3(u^2)_{x},\\
\label{phi4}
(\phi^4)\ & \phi_{xx} -\phi _{tt} &=m^{2}\phi - \lambda\phi^{3},\\
\label{sg}
({\rm SG})\ &\phi_{xx} - \phi_{tt}&=\frac{m^{3}}{\sqrt{\lambda}}
sin[(\frac{\sqrt{\lambda}}{m})\phi].
\end{eqnarray} 

The above models have quite different intrinsic characteristics.
First of all SG is an integrable models while BBMP and $\phi^4$
are not. Second the linear dispersion relation $\omega{(k)}$  
has a finite limit as  $k\to\infty$ (SW limit) for BBMP ,
while it is unbounded for SG and $\phi^4$. Indeed we have
\begin{eqnarray}
\omega_{(BBMP)} &=&\frac{k}{1 + k^{2}},
\label{3}\\
\omega_{(\phi^{4})} &=&\omega_{(SG)}=(m^{2} + 
k^{2})^\frac12,\label{6}
\end{eqnarray}

However the phase and group velocities are all bounded in the SW limit
$k\to\infty$, which is a central point in this approach as indeed
this very property allows the three models to sustain short waves.
Then we face the problem of the {\em nonlinear
propagation of a SW}, which is the object of this work. In the following,
both nonintegrable systems (BBMP and $\phi^4$) will be displayed in
details, while the integrable one (SG) will be only sketched.

\paragraph*{The BBMP Model.}

Let us consider a
SW in (1) characterized by $k={k_{0}}{\epsilon^{-1}}$ with 
$k_{0}\sim {\cal O}(1)$ and $\epsilon\ll1$. The plane wave solution of the
linear problem $u = \exp\,i\{kx - \omega(k)t\}$ inspires
a fast variable $\zeta =\epsilon^{-1}x$ and infinitely many
slow time variables $\tau_{2n + 1} = \epsilon^{2n + 1}t$ 
($n = 0, 1, 2, ...$), by expanding  $\omega$ in powers of $\epsilon$.

We assume the expansion
\begin{equation}
u = u_{0} + \epsilon^{2}u_{2} + \epsilon^{4} u_{4} + ...\,.
\label{7}
\end{equation}
and suppose the {\em extension} 
$u_{2n} = u_{2n}(\zeta,\tau_{1},\tau_{3},\cdots)$, $n = 0,1,\cdots,$ 
\cite{jeffrey,sandri}. Then, the operators
\begin{eqnarray}
\frac{\partial}{\partial x} 
&=&\frac1\epsilon \frac{\partial}{\partial\zeta},
\label{6a}\\
\frac{\partial}{\partial t} 
&=& \epsilon \frac{\partial}{\partial\tau_{1}}
+\epsilon^{3}\frac{\partial}{\partial\tau_{3}} 
+\epsilon^{5}\frac{\partial}{\partial\tau_{5}} + ...\,.
\label{6b}
\end{eqnarray}
allow us to study the behavior of a {\em short-wave} for {\em large
 time}.

BBMP gives at orders  $\epsilon^{-1}, 
\epsilon, \epsilon^{3}$,
..., the equations (written only up $\epsilon^{3}$)
\begin{eqnarray}
&&-u_{0,\zeta\tau_{1}} + u_{0} -
3u_{0}^{2} = 0,
\label{8}\\
&&\hat Lu_{2} = u_{0,\tau_{1}} + 
u_{0,\zeta\zeta\tau_{3}},  
\label{9}\\
&&\hat Lu_{4} = u_{2,\tau_{1}}
 - u_{0,\tau_{3}} + 
u_{2,\zeta\zeta\tau_{3}} + 
u_{0,\zeta\zeta\tau_{5}} +
3(u_{2}^{2})_{\zeta},
\label{10}\end{eqnarray}
where $\hat L$ is the linear operator associated with (\ref{8}): 
\begin{equation}
\hat L(\upsilon) = -\upsilon_{\zeta\zeta\tau_{1}} + \upsilon_{\zeta}
 - 6(\upsilon u_{0})_{\zeta}.
\label{10bis}
\end{equation}

The {\em unique} solution of (\ref{8}) in the
form $ u_{0}(\eta)$ with $\eta = k_{0}\zeta - \omega_{1}\tau_{1} -
\omega_{3}\tau_{3} - \omega_{5}\tau_{5} ...,$\, going to zero for
 $\left|\zeta\right|\rightarrow\infty$ is 
\begin{equation}
u_{0} = \frac12{\rm sech}^{2}\eta, \,\,\,\, \omega_{1} =
-\frac{1}{4k_{0}}.
\label{11}
\end{equation}
The values $\omega_{3}, \omega_{5},\cdots$, corrections to the principal
frequency $\omega_{1}$ (Stokes' hypoyhesis) are still free but will be
determined later by the non-secularity requirement.

The equation (\ref{9}) for $u_{2}$ then reads: 
\begin{equation}
\hat Lu_{2} = \{4\omega_3 k_0^2 - \omega_1
 - 12\omega_3 k_0^2{\rm sech}^2\eta\}{\rm sech}^2\eta\tanh\eta,
\label{12}
\end{equation}
and its two first right hand side terms are resonant (secular
producing terms) because \cite{nayfeh} 
$$
\hat L({\rm sech}^{2}\eta{\rm
tanh}\eta)= 0.
$$
These secular terms are eliminated by choosing 
$$\omega_{3} =\frac{\omega_1}{4k_0^2}=-\frac1{4^2k_0^3}.$$
Hence, (\ref{12}) yields the solution $u_{2}(\eta) =
4^{-1}k_{0}^{-2}u_{0}(\eta)$. 

The equation (\ref{10}) for $u_{4}(\eta)$ contains secular
producing terms originated by the first four terms in the right hand
side. They can be eliminated choosing $\omega_{5} =
-4^{-3}k_{0}^{-5}$. The solution is $u_{4}(\eta) =
4^{-2}k_{0}^{-4}u_{0}(\eta)$.
This procedure can be repeated at any higher order $n = 0,1,2,..$  and 
we obtain recursively
\begin{equation}
u_{2n}(\eta) = \frac{u_{0}(\eta)}{4^{n}k_{0}^{2n}},\quad
 \omega_{2n + 1} = -\frac1{4^{n + 1} k_0^{2n + 1}}.  
\label{13}\end{equation}

Next, nicely enough, not only the perturbative series solution (\ref{7})
can be  summed to give
\begin{equation}
u(\eta) = u_{0}(\eta)\sum_{n=0}^{\infty}\frac{\epsilon^{2n}}{4^{n}
k_{0}^{2n}} =\frac{4k^{2}}{4k^{2} - 1}u_{0}(\eta).
\label{14}
\end{equation}
but also, by using $\omega_{2n + 1}$, its argument $\eta$ in the 
laboratory coordinates, which results as
\begin{equation}
\eta = kx + \frac{1}{4k}\sum_{n=0}^{\infty}\frac{t}{(4k^{2})^{n}} 
= kx + \frac{kt}{4k^{2}- 1}.
\label{15}
\end{equation}

Therefore, this SW perturbation technique finally leads to the solution 
\begin{equation}
u(x,t) = -\frac{2k^{2}}{1 - 4k^{2}}{\rm sech}^{2}[k(x +
\frac{t}{4k^{2} - 1})].
\label{16}
\end{equation}
This very expression, solution of BBMP,
was obtained in \cite{kraenkel1} as an
asymptotic limit of a LW of small amplitude. 
Thus, for $t\rightarrow\infty$, the nonlinear dynamics of a SW
(with an order one amplitude) and that of a LW (with small amplitude) are
indistinguishable in BBMP. The equation (\ref{8})
 $$u_{0,\zeta\tau_{1}} = u_{0} -3u_{0}^{2}$$
is a nonlinear Klein-Gordon
equation which sustitutes in this SW approach the classical
Korteweg-de Vries of the LW approach.

\paragraph*{ The $\phi^{4}$ Model.}

The topological antikink type solution of $\phi^{4}$ will be obtained by
perturbation expansion starting from the constant solutions $\phi_{0} = 
\pm{m}/{\sqrt{\lambda}}$. Hence we seek a solution $\phi(\eta)$ 
such that $\phi\to
\mp{m}/{\sqrt{\lambda}}$ for $\eta\to\pm\infty$.

For $\eta < 0$, the function $u = \phi - 
{m}/{\sqrt{\lambda}}$, 
goes to zero for $\eta\to-\infty$ and satifies 
\begin{equation}
u_{xx} - u_{tt} = -2m^{2}u - 3m\sqrt{\lambda}u^{2} - \lambda u^{3}.
\label{23}
\end{equation}
For unidirectional propagation the convenient fast variable
$\zeta$ and the slow variables $\tau_{2n + 1}$ are in this case: 
$\zeta = \epsilon^{-1} (x - t)$,
$\tau_{1} = \epsilon t, \tau_{3} = \epsilon^{3} t, \cdots$. Expanding
$u$ according to $u = \epsilon^{2}(u_{0} + \epsilon^{2}u_{2} +
\epsilon^{4}u_{4} +\cdots)$,  the 
 resulting equations  are (up to $\epsilon^{6}$),
\begin{eqnarray}
\hat Lu_{0}& =& 0 ,
\label{u-0}  \\
 \hat Lu_{2} &=& -2u_{0,\zeta\tau_{3}} + 
u_{0,2\tau_{1}} - 3m \sqrt{\lambda} u_{0}^{2}, 
\label{24}  \\
\hat Lu_{4} &=& u_{2,2\tau_{1}} -2u_{2,\zeta\tau_{3}}  
 - 2u_{0,\zeta\tau_{5}} + 
2u_{0,\tau_{1}\tau_{3}} \nonumber \\
& & - 6m\sqrt{\lambda}u_{0}^{2}u_{2} - \lambda
u_{0}^{3}. 
\label{25} 
\end{eqnarray}
with $\hat L $ in this case being the linear Klein-Gordon operator
\begin{equation}
\hat L(\upsilon) = 2 \upsilon_{\zeta\tau_{1}} + 
2m^{2} \upsilon
\label{25bis}\end{equation}

We choose for the solution $u_{0}$ of (\ref{u-0})
the form $u_{0} = B\exp2\eta$ with $\eta =
k_{0}\zeta - (4k_{0})^{-1}m^{2}\tau_{1} + \omega_{3}\tau_{3} +
\cdots$ and $B$  a constant. All linear terms at the right hand
side of equations for $u_{2(n - 1)}$ are secular. They can be
eliminated choosing appropiately $\omega_{2n - 1}$, namely
\begin{equation}
\omega_{2n - 1} = -\frac{(\frac{1}{2})!}{n!(\frac{1}{2} -
n)!}\frac{m^{2n}}{2^{n}k_{0}^{2n - 1}}.
\label{26}
\end{equation}
Next the solutions read
\begin{equation}
u_{2(n-1)} = B^{n}(\frac{\sqrt{\lambda}}{2m})^{n - 1}\exp2n\eta.
\label{27}
\end{equation}

With these values of $\omega_{2n - 1}$ the series for $\eta$
can again   be   summed as
$$\eta = kx - \sqrt{ k^{2} +\frac{m^{2}}{2}}\ t,$$ 
and also the perturbative series for $u$ only if we choose
$B = - \frac{2m}{\sqrt{\lambda}}k_{0}^{-2}$. It leads for
$\phi=u+m/\sqrt{\lambda}$ to
\begin{equation}
\phi = -\frac{m}{\sqrt{\lambda}} \tanh \{ kx - \sqrt{ k^{2} + 
\frac{m^{2}}{2}}\,\,t - \log k \},
\label{29}
\end{equation}
To get this expression it is necessary to use
the Fourier representation $(x<0)$
$$
\sum_{n=0}^{\infty}(-1)^{n +1}\delta_{n} \exp(2nx) = \tanh x,
$$
where $\delta_{n}$ are the Neumann's numbers ($\delta_{0} = 
1,\,\, \delta_{n} = 2,\,\, \forall\,\, n = 1,2,3, ... $).

The above solution $\phi$ is the antikink solution of  $\phi^{4}$ 
(with an initial shift $\log k/k$), which has not been obtained
previously within another perturbation scheme. 

The expression $\sqrt{k^{2} +\frac{m^{2}}{2}}$ can be interpreted
as a nonlinear frequency $\omega_{nl}$, which defines the nonlinear
group velocity 
\begin{equation}
v = \frac{\partial \omega_{nl}}{\partial k} = \frac{k}{ 
\sqrt{k^{2} + \frac{m^{2}}{2}}}.
\label{31}
\end{equation}
It is remarkable that the Lorentz invariance of (\ref{29}) is 
precisely related to that particular velocity, indeed
\begin{equation}
\phi = -\frac{m}{\sqrt{\lambda}}\tanh \{
\frac{m}{\sqrt{2}}(\frac{xv}{\sqrt{1 - v^{2}}} - 
\frac{t}{\sqrt{1 - v^{2}}}) - \log k \}.
\label{32}
\end{equation}

Note that the case $\eta > 0 $ in the perturbative series would simply
yield the solution  $\phi(-\eta)$.

\paragraph*{The SG Model.}

Finally in the case of the sine-Gordon model (\ref{sg}),  
for $\phi = \epsilon( \phi_{0} +
 \epsilon^{2}\phi_{2} +
\epsilon^{4}\phi_{4} + ...\,\,\,)$, with $\phi_{2n}$ functions of 
$\eta = k_{0}\zeta + (2k_{0})^{-1}m^{2}\tau_{1} + \omega_{3}\tau_{3} +
 ...$ where $\zeta = \epsilon^{-1} (x - t)$,
$\tau_{1} = \epsilon t, \tau_{3} = \epsilon^{3} t, \cdots$,  
we obtain (up to order $\epsilon^{4}$)

\begin{eqnarray}
\hat L(\phi_{0})&=& 0, 
\label{32a} \\
\hat L(\phi_{2}) &=& -2\phi_{0, \zeta\tau_{3}} + \phi_{0, 2\tau_{1}} - 
\frac{\lambda}{3!}\phi_{0}^{3}, 
\label{32b} \\
\hat L(\phi_{4}) &=& -2\phi_{2, \zeta\tau_{3}} + \phi_{2, 2\tau_{1}}  
 -2\phi_{0, \zeta\tau_{3}} + 2\phi_{0, \tau_{1}\tau_{3}}\nonumber \\
& &-\frac{\lambda}{3!}3\phi_{0}^{2}\phi_{2} +
\frac{\lambda^{2}}{m^{2}}\frac{\phi_{0}^{5}}{5!},  
\end{eqnarray}

with $\hat L$ being in this case the operator

\begin{equation}
\hat L(\upsilon) = 2\upsilon_{\zeta\tau_{1}} - m^{2}\upsilon.
\label{32d}
\end{equation}
We choose for the solution of (\ref{32d}) the expression $\phi_{0} =
C\exp\eta $ with $C$ a constant. As in the previous case all the linear
terms at the right hand side of equations for $\phi_{2(n - 1)}$ are
secular. They can be eliminated choosing $\omega_{2(n-1)}$ as

\begin{equation}
\omega_{2n - 1} = (-1)^{n + 1}\frac{(\frac{1}{2})!}{n!(\frac{1}{2} -
n)!}\frac{m^{2n}}{k_{0}^{2n - 1}}.
\label{33}
\end{equation}
Hence the solutions $\phi_{2(n - 1)}$ read
\begin{equation}
\phi_{2(n-1)} = -\frac{-\lambda}{16})^{n - 1}\frac{C^{2n - 1}\exp(2n -1)
 \eta}{m^{2(n - 1)}(2n - 1) }.
\label{33a}
\end{equation}
The series for $\phi$ summs for $C = 4m/{\sqrt{\lambda}k_{0}}$ 
and yields 
\begin{eqnarray}
\phi &=& \frac{4m}{\sqrt{\lambda}}\sum_{n=0}^{\infty}
(-1)^{n}\frac{\exp[(2n +1)(\eta - \log k)]}{2n +1} ,\nonumber\\
&=&\frac{4m}{\sqrt{\lambda}}\arctan \left\{ \exp(kx - \sqrt{k^{2} -
m^{2}}\,\,t - \log k)\right\}.
\label{34}
\end{eqnarray}

In this case the Lorentz invariant form of (\ref{34}) appears as
a function of the nonlinear phase velocity 
$v =  \sqrt{1 - m^2/k^2}$ 
as
\begin{equation}
\phi = \frac{4m}{\sqrt{\lambda}}\arctan \left\{ \exp \left[ 
\frac{m}{\sqrt{1 - v^{2}}}( x - vt) - \log k \right] \right\}.
\label{35}
\end{equation}

\paragraph*{Conclusion and Comments}

We have applied a multiple-time version of the reductive perturbation
method to study the solitary-wave and the kink-wave solutions of some
nonlinear dispersive models. These solutions has already been known
before. The alternative way gives here to obtain them shows that they
represented {\it short-waves asymptotic dynamics}
($t\rightarrow\infty$)\\

1 - BBMP for long-waves serves about the same purpose as KdV, whereas their
behaviors in propagating short-waves can be expected to be rather
different from KdV: from a linear analysis,
BBMP does not propagate short-waves while KdV amplify
them\cite{Bona}. Then our result answers  this old
controversy on the relative relevance between KdV and BBMP. 
\cite{Kruskal}. Actually we proved that
short-waves do propagate nonlinearly in BBMP models,
and build up soliton-like solutions as $t\rightarrow\infty$.

2 - The antikink (or kink) solution of $\phi^{4}$ model which cannot
be obtained as a perturbative solution in $\lambda$ \cite{rajara}, 
appears as a perturbative solution in $k$ in the short-wave limit.

3 - The eq. (\ref{34}) shows that the kink solution of SG is
obtainable {\it only} from a short-wave dynammics, as indeed, the limit $k\to0$
gives rise to an imaginary argument.

4 - An initial profile generically contains short-wave components which
are usually neglected in favor of the long-wave components, which
occurs as well
through numerical discretisation of the models. As we have shown that the
short-wave components asymptotically build up soliton solutions, the 
common understanding of a soliton as originating from long-wave is to be
questionned.

5 - It is worth noting finally that, within the long-wave approach, 
the nonlinear character of the solution is present
already at the first orders, as
indeed one usually finds the Boussinesq, KdV, MKdV, etc equations. This is
not the
case with the short-wave approach where usually {\em all orders} (i.e. all
times)
are necessary to unveil the nonlinear character of the solution.

\paragraph*{Aknoledgements}

The authors wish to thank J. Leon and P. Grang\'e for
many helpful and stimulating discussion. One of us (M.A.M) is indebted
to  R. A. Kraenkel and J. G. Pereira for fruitfull
collaboration.

\end{multicols}

\end{document}